\documentclass[sigconf, nonacm]{acmart}

\usepackage{acmart-taps}
\usepackage{xspace}
\usepackage{balance}
\usepackage{graphics}
\usepackage{color}
\usepackage{todonotes}
\usepackage{xspace}
\usepackage{siunitx}
\usepackage[flushleft]{threeparttable}
\usepackage{enumitem}
\usepackage{balance}
\usepackage{caption}
\usepackage{graphics}
\usepackage{color, colortbl, soul, xcolor}
\usepackage{tikz}
\usepackage{caption,subcaption}
\usepackage{fixltx2e}
\usepackage{bm}
\usepackage{romannum}
\usepackage{tabularx,booktabs,caption,ragged2e}
\usepackage[title]{appendix}
\usepackage{color, colortbl, soul, xcolor}
\usepackage{bm}
\usepackage{algorithm}
\usepackage{algorithmic}
\usepackage[utf8]{inputenc}
\usepackage{tcolorbox}
\usepackage{hyperref}
\usepackage{changepage}
\usepackage[normalem]{ulem}

\def\etal{\textit{et~al.}\xspace}
\def\eg{\textit{e.g.,}\xspace}

\begin{document}

\def\sysname{\mbox{RemiVoice}}

\author{Aaryan Gajula}
\authornote{Both authors contributed equally to this paper.}
\orcid{0009-0002-2846-2007}
\email{agaju003@fiu.edu}
\affiliation{%
  \institution{Florida International University}
  \city{Miami}
  \state{FL}
  \country{USA}
}

\author{Soumay Agarwal}
\authornotemark[1]
\orcid{0009-0009-4127-5610}
\email{soa002@ucsd.edu}
\affiliation{%
  \institution{University of California San Diego}
  \city{La Jolla}
  \state{CA}
  \country{USA}
}

\author{Shaoze Zhou}
\orcid{0009-0000-3243-0599}
\email{szhou010@fiu.edu}
\affiliation{%
  \institution{Florida International University}
  \city{Miami}
  \state{FL}
  \country{USA}
}

\author{Lingyao Li}
\orcid{0000-0001-5888-8311}
\email{lingyaoli@arizona.edu}
\affiliation{%
  \institution{University of Arizona}
  \city{Tucson}
  \state{AZ}
  \country{USA}
}

\author{Renkai Ma}
\orcid{0000-0002-4434-2235}
\email{mark@ucmail.uc.edu}
\affiliation{%
  \institution{University of Cincinnati}
  \city{Cincinnati}
  \state{OH}
  \country{USA}
}

\author{Jennifer Martin}
\orcid{0000-0003-0849-3391}
\email{jlmartin@fiu.edu}
\affiliation{%
  \institution{Florida International University}
  \institution{Miami VA Health Care System}
  \city{Miami}
  \state{FL}
  \country{USA}
}

\author{Krisstina Madan}
\orcid{0009-0008-1229-5105}
\email{kmadan@fiu.edu}
\affiliation{%
  \institution{Florida International University}
  \city{Miami}
  \state{FL}
  \country{USA}
}

\author{Ellen Brown}
\orcid{0000-0002-2418-3257}
\email{ebrown@fiu.edu}
\affiliation{%
  \institution{Florida International University}
  \city{Miami}
  \state{FL}
  \country{USA}
}

\author{Chen Chen}
\orcid{0000-0001-7179-0861}
\email{chechen@fiu.edu}
\affiliation{%
  \institution{Florida International University}
  \city{Miami}
  \state{FL}
  \country{USA}
}

\renewcommand{\shortauthors}{Gajula and Agarwal~\etal}

\keywords{Reminiscence Therapy, Older Adults, Dementia, Conversational Voice User Interface, Generative AI}

\title[\sysname]{\sysname: Supporting Reminiscence Therapy for Older Adults with Mild Dementia Through Voice-First Conversational AI}

\begin{teaserfigure}
    \centering
    \includegraphics[width=\linewidth]{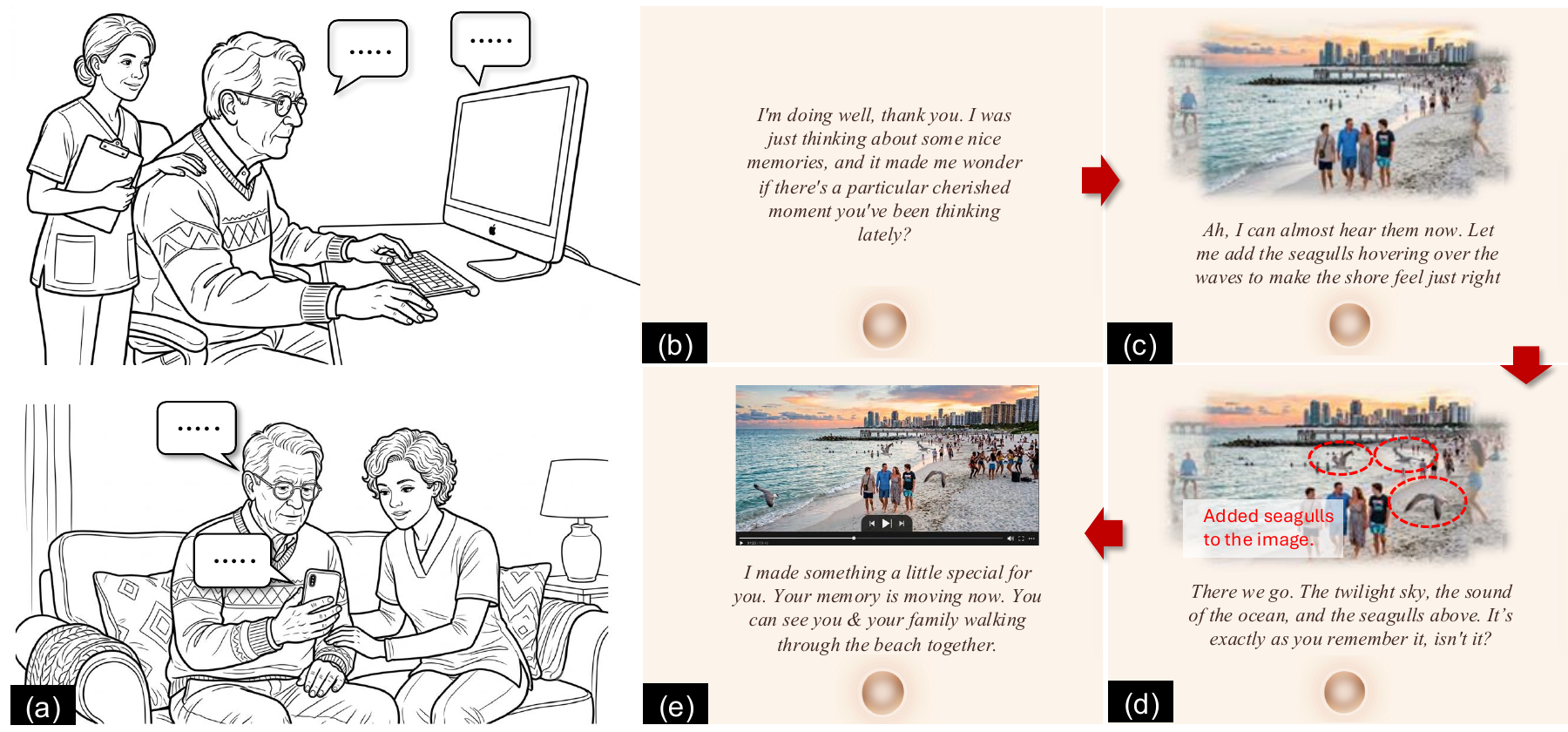}
    \vspace{-.25in}
    \caption{(a) \sysname~can be accessed on both desktop computers and mobile devices through a voice-first interaction experience, which begins by collecting memory cues through conversation (b), generates personalized images from users' verbal descriptions (c), enables iterative refinement of images via voice-first interactions (d), and finally creates AI-generated videos to help users visualize and reconstruct their memories (e).}
    \label{fig::teaser}
\end{teaserfigure}

\begin{abstract}

With the global population aging and increasing prevalence of dementia, there is an urgent need for effective solutions to support patients across various stages of \textbf{A}lzheimer's \textbf{D}isease and \textbf{R}elated \textbf{D}ementias~(ADRD).
\textbf{R}eminiscence \textbf{T}herapy (RT) is a validated intervention designed to trigger memories and is widely used for
various stages of dementia. 
We present our preliminary prototype and exploration of \emph{\sysname}, a browser-based voice-first conversational AI assistant that supports older adults with mild dementia in RT through conversational grounded images and videos.

\end{abstract}

\maketitle

\section{Introduction}

With the global population aging and increasing prevalence of dementia, there is an urgent need for effective solutions to support patients across various stages of \textbf{A}lzheimer's \textbf{D}isease and \textbf{R}elated \textbf{D}ementias (ADRD)~\cite{Altizer2025, Pond2026}.
\textbf{R}eminiscence \textbf{T}herapy (RT) is a validated intervention designed to trigger memories and is widely used for various stages of dementia~\cite{RT2026}.
While reminiscence therapy (RT) is a validated intervention for individuals with dementia, traditional RT often relies on personal photographs or physical artifacts, which many patients with ADRD~\cite{Altizer2025, Pond2026} may lack or have difficulty accessing. This limitation creates a significant barrier for healthcare professionals and caregivers seeking to facilitate meaningful conversations about past experiences that promote comfort, enjoyment, and cognitive stimulation.

Recent digital technologies enable a scalable and cost-effective solution to support RT for older adults~\cite{Zhang2026RTSurvey, InspireD2026, Boyd2021, Lazar2014, Kuwahara2006}. 
For example, Remme~\cite{Remme} enables patients to engage in natural voice-based conversations with a virtual therapist centered on the imported photographs.
Similar AI-guided RT conversations grounded in personal props have also been widely explored~\cite{Seah2026Rememo, Nebot2022LONGREMI, Wang2024GoodTimes, Astell2010, Jiang2025Remini, Mariona2020}.
\textbf{M}ultimodal \textbf{L}arge \textbf{L}anguage \textbf{M}odels~(MLLMs) have expanded opportunities for facilitating RT without relying on personal photographs.
\mbox{RemiHaven}~\cite{Zhang2025RemiHaven} introduced a text-based interface that enables older migrants to conduct self-managed RT by creating and editing images based on their residual memories.
Nan~\etal~\cite{Nan2025KimonoEra} explored a similar chat-based interface that enables older adults to conduct RT conversations using uploaded images and stylistically transformed versions of them.
Reminiscope~\cite{Zhu2026Reminiscope} leveraged AI to transform handcrafted textile collages into memory-related video content in virtual reality.
ReminiBuddy~\cite{Sun2025ReminiBuddy} explored AI-guided RT conversations grounded in AI-generated images and 3D objects, facilitated by two conversational agents with distinct identities.
Robo-Blocks~\cite{Kim2025RoboBlocks} explored the feasibility of using an LLM-enabled chat interface to support intergenerational co-creation activities.

To overcome the cognitive and physical burdens associated with text-based interfaces (\eg \cite{Zhang2025RemiHaven, Nan2025KimonoEra, Jiang2025Remini}), we present our work-in-progress prototype of \emph{\sysname}, a voice-first conversational AI assistant that assists older adults with mild dementia in conducting RT through AI-generated images and videos, without requiring personal photographs.
\sysname~can be accessed through a web application on desktop or mobile devices (Figure~\ref{fig::teaser}a). 
While patients can use RemiVoice independently, we recommend that a caregiver accompany the patient during RT conversation to help prevent and respond to potentially unexpected or distressing moments.
\sysname~begins the RT experience by collecting memory cues through natural voice-based conversations (Figure~\ref{fig::teaser}b), generating and iteratively co-editing images with the patients (Figure~\ref{fig::teaser}c - d), and ultimately creating videos that help older adults visualize and revisit their memories (Figure~\ref{fig::teaser}e).

\section{Design and Prototyping~\sysname}

\noindent{\bf Voice-first RT experience.}
We design \sysname~ as a voice-first user experience that does not require users to input commands or information via keyboard and/or mouse. 
The voice-first interaction paradigm prioritizes conversational speech as the primary mode of interaction, while using visual elements as secondary supports~\cite{Whitenton2017VoiceFirst, Murad2022}. 
This approach has been successfully implemented in designing voice-based digital tools for older adults~\cite{Chen2023ScreenOrNoScreen, Chen2021, Han2022, Chen2023Setup, Cuadra2022, Lifset2023}.
Grounded in a specific task, the AI agent begins by eliciting key \textbf{memory cues}, including the relevant \emph{time period}, \emph{environment}, and \emph{people} associated with the recalled experience.
Based on the collected memory cues, the agent generates multiple relevant scenes that represent the patient's recollection. 
We generate five images from different camera perspectives - a wide establishing shot, an extreme close-up of one named detail, a first-person view, a low angle, and a bird's-eye view - to offer different retrieval cues.
The patient can then iteratively co-refine the generated images with the AI through natural conversations while continuing the RT conversation.
Finally, a video is generated from the refined scene and incorporated into an interactive digital storybook, allowing the patient to revisit it during future RT sessions.

\vspace{2px}\noindent{\bf A multi-agent system design.}
\sysname~was implemented as a multi-agent system consisting of three agentic pipelines;
the \textbf{conversational agent}, orchestrated through a dedicate LangGraph pipeline~\cite{langgraph}, facilitates empathetic reminiscence conversations with older adult patients. The \textbf{image agent} generates and iteratively edits AI-created images based on the patient's contextual memory cues;
the \textbf{video agent} synthesizes personalized videos from the refined images and the collected contextual information.
Our AI agents were implemented by Gemini Live~\cite{GeminiLive}, Gemini 2.5 Flash Image~\cite{GeminiFlashImage}, and Gemini Veo 3.1~\cite{GeminiVeo}, respectively.

\section{Early Design Reflections}
%

We adopted an autobiographical approach~\cite{Neustaedter2012, Lucero2019, Neustaedter2012b, Brown2023} to reflect our work-in-progress prototype within our research team. Our team includes researchers from Computer Science, Geriatrics, and Gerontology.
The work-in-progress implementation of \sysname~served as a prototype~\cite{Mogensen1992, Boer2012} to facilitate our discussion.
Team members then shared insights informed by their interdisciplinary expertise.
Our initial reflection led to three key findings:

\vspace{2px}\noindent{\bf Value of voice-first conversational experience.}
Geriatric experts appreciated the value of the voice-first conversational experience, particularly the empathetic and caring voice of the simulated agent as well as the simple, neutral, and consistent visual interface, despite using voice as the primary modality for interaction.

\vspace{2px}\noindent{\bf Safety and emotion awareness.}
Our discussion raised several safety concerns that are also broadly generalizable to the use of AI for therapeutic purposes~\cite{Li2026}.
While appreciated the AI-generated images, our discussion highlighted two primary concerns. 
First, AI-generated images may evoke unexpected negative emotions, such as fear or distress.
For example, one geriatric expert with home care experience commented: {\color{gray} \it ``there is an image when giant people standing on the taxis [...] that may potentially frighten people living with dementia [...] We don't know whether AI will generate inappropriate content? Maybe this person has a phobia of spiders, and there is a spider on there in the image.''}~
Second, our discussion highlighted the critical need to monitor users' emotional states and adopt adaptive strategies when conversations become emotionally negative or unintentionally trigger stress and distress.
We also highlighted the importance of having a caregiver accompany the patient while using \sysname, rather than having the patient engage in reminiscence therapy conversations alone.

\vspace{2px}\noindent{\bf Visual stimuli beyond images.}
While static images were considered useful, we highlighted the potential value of incorporating additional stimuli beyond visual content. Although geriatric experts noted that videos may overstimulate users with dementia, they suggested that audio-based stimuli, such as music and ambient sounds, could be valuable additions to the experience.
\section{Future Work}
Our reflection reveals three future directions.
\emph{First}, we will investigate how to design and adapt RT conversations based on patients' cognitive and emotional states. 
Recalling certain memories may also evoke distressing emotions or emotional flooding~\cite{Kelly2021, Khan2022}. 
Future iterations of \sysname~will explore adaptive conversation-steering strategies that dynamically redirect or reframe the dialogue when signs of emotional distress are detected.
A \emph{second} direction is to evaluate the usability of \sysname~and its effectiveness in supporting RT through longitudinal deployment studies involving older adults with mild dementia.
\emph{Finally}, we will investigate the effectiveness of \sysname~among more specific aging populations, such as older veterans and older migrants, whose life experiences, cultural backgrounds, and reminiscence needs may differ substantially.

\begin{acks}

This work was partially supported by the startup grant provided by \textbf{F}lorida \textbf{I}nternational \textbf{U}niversity (FIU) and the research credit award from Google.
We thank the valuable feedback and comments from our colleagues at the Knight Foundation School of Computing and Information Sciences at FIU.

\end{acks}

\balance
\bibliographystyle{ACM-Reference-Format}
\bibliography{reference}

@article{Lifset2023,
    title={Ascertaining whether an intelligent voice assistant can meet older adults’ health-related needs in the context of a geriatrics 5Ms framework},
    author={Lifset, Ella T and Charles, Kemeberly and Farcas, Emilia and Weibel, Nadir and Hogarth, Michael and Chen, Chen and Johnson, Janet G and Draper, Mary and Nguyen, Annie L and Moore, Alison A},
    journal={Gerontology and Geriatric Medicine},
    volume={9},
    pages={23337214231201138},
    year={2023},
    publisher={SAGE Publications Sage CA: Los Angeles, CA},
    doi={10.1177/23337214231201138}
}

@article{Jiang2025Remini,
    author = {Jiang, Zhuoqun and Yeo, ShunYi and Seow, Wei Xuan, Donovan and Perrault, Simon Tangi},
    title = {Remini: Leveraging Chatbot-Mediated Mutual Reminiscence for Promoting Positive Affect and Feeling of Connectedness among Loved Ones},
    year = {2025},
    issue_date = {November 2025},
    publisher = {Association for Computing Machinery},
    address = {New York, NY, USA},
    volume = {9},
    number = {7},
    url = {https://doi.org/10.1145/3757650},
    doi = {10.1145/3757650},
    journal = {Proc. ACM Hum.-Comput. Interact.},
    month = oct,
    articleno = {CSCW469},
    numpages = {43}
}

@inproceedings{Mariona2020,
    author = {Car\'{o}s, Mariona and Garolera, Maite and Radeva, Petia and Giro-i-Nieto, Xavier},
    title = {Automatic Reminiscence Therapy for Dementia},
    year = {2020},
    isbn = {9781450370875},
    publisher = {Association for Computing Machinery},
    address = {New York, NY, USA},
    url = {https://doi.org/10.1145/3372278.3391927},
    doi = {10.1145/3372278.3391927},
    booktitle = {Proceedings of the 2020 International Conference on Multimedia Retrieval},
    pages = {383–387},
    numpages = {5},
    location = {Dublin, Ireland},
    series = {ICMR '20}
}

@inproceedings{Neustaedter2012,
    author = {Neustaedter, Carman and Sengers, Phoebe},
    title = {Autobiographical design in HCI research: designing and learning through use-it-yourself},
    year = {2012},
    isbn = {9781450312103},
    publisher = {Association for Computing Machinery},
    address = {New York, NY, USA},
    url = {https://doi.org/10.1145/2317956.2318034},
    doi = {10.1145/2317956.2318034},
    booktitle = {Proceedings of the Designing Interactive Systems Conference},
    pages = {514–523},
    numpages = {10},
    location = {Newcastle Upon Tyne, United Kingdom},
    series = {DIS '12}
}

@inproceedings{Lucero2019,
    author = {Lucero, Andr\'{e}s and Desjardins, Audrey and Neustaedter, Carman and H\"{o}\"{o}k, Kristina and Hassenzahl, Marc and Cecchinato, Marta E.},
    title = {A Sample of One: First-Person Research Methods in HCI},
    year = {2019},
    isbn = {9781450362702},
    publisher = {Association for Computing Machinery},
    address = {New York, NY, USA},
    url = {https://doi.org/10.1145/3301019.3319996},
    doi = {10.1145/3301019.3319996},
    booktitle = {Companion Publication of the 2019 on Designing Interactive Systems Conference 2019 Companion},
    pages = {385–388},
    numpages = {4},
    location = {San Diego, CA, USA},
    series = {DIS '19 Companion}
}

@article{Neustaedter2012b,
    author = {Neustaedter, Carman and Sengers, Phoebe},
    title = {Autobiographical design: what you can learn from designing for yourself},
    year = {2012},
    issue_date = {November + December 2012},
    publisher = {Association for Computing Machinery},
    address = {New York, NY, USA},
    volume = {19},
    number = {6},
    issn = {1072-5520},
    url = {https://doi.org/10.1145/2377783.2377791},
    doi = {10.1145/2377783.2377791},
    journal = {Interactions},
    month = {nov},
    pages = {28–33},
    numpages = {6}
}

@inproceedings{Li2026,
    author = {Li, Lingyao and Huang, Xiaoshan and Ma, Renkai and Zhang, Ben Zefeng and Wu, Haolun and Yang, Fan and Chen, Chen},
    title = {LLM Use for Mental Health: Crowdsourcing Users' Sentiment-based Perspectives and Values from Social Discussions},
    year = {2026},
    isbn = {9798400723070},
    publisher = {Association for Computing Machinery},
    address = {New York, NY, USA},
    url = {https://doi.org/10.1145/3774904.3793025},
    doi = {10.1145/3774904.3793025},
    booktitle = {Proceedings of the ACM Web Conference 2026},
    pages = {9687–9698},
    numpages = {12},
    location = {United Arab Emirates},
    series = {WWW '26}
}

@article{Mogensen1992,
    title={Towards a Provotyping Approach in Systems Development.},
    author={Mogensen, Preben},
    journal={Scand. J. Inf. Syst.},
    volume={4},
    number={1},
    pages={5},
    year={1992}
}

@article{Brown2023,
    title={Developing a memory and communication app for persons living with dementia: an 8-step process},
    author={Brown, Ellen L and Ruggiano, Nicole and Allala, Sai Chaithra and Clarke, Peter J and Davis, Debra and Roberts, Lisa and Framil, C Victoria and Mu{\~n}oz, Mar{\'\i}ateresa Teri Hernandez and Hough, Monica Strauss and Bourgeois, Michelle S},
    journal={JMIR aging},
    volume={6},
    pages={e44007},
    year={2023},
    publisher={JMIR Publications},
    doi={10.2196/44007}
}

@inproceedings{Boer2012,
    author = {Boer, Laurens and Donovan, Jared},
    title = {Provotypes for participatory innovation},
    year = {2012},
    isbn = {9781450312103},
    publisher = {Association for Computing Machinery},
    address = {New York, NY, USA},
    url = {https://doi.org/10.1145/2317956.2318014},
    doi = {10.1145/2317956.2318014},
    booktitle = {Proceedings of the Designing Interactive Systems Conference},
    pages = {388–397},
    numpages = {10},
    location = {Newcastle Upon Tyne, United Kingdom},
    series = {DIS '12}
}

@article{Lazar2014,
    title={A systematic review of the use of technology for reminiscence therapy},
    author={Lazar, Amanda and Thompson, Hilaire and Demiris, George},
    journal={Health education \& behavior},
    volume={41},
    number={1\_suppl},
    pages={51S--61S},
    year={2014},
    publisher={SAGE Publications Sage CA: Los Angeles, CA},
    doi={10.1177/1090198114537067}
}

@inproceedings{Kuwahara2006,
    author = {Kuwahara, Noriaki and Abe, Shinji and Yasuda, Kiyoshi and Kuwabara, Kazuhiro},
    title = {Networked reminiscence therapy for individuals with dementia by using photo and video sharing},
    year = {2006},
    isbn = {1595932909},
    publisher = {Association for Computing Machinery},
    address = {New York, NY, USA},
    url = {https://doi.org/10.1145/1168987.1169010},
    doi = {10.1145/1168987.1169010},
    booktitle = {Proceedings of the 8th International ACM SIGACCESS Conference on Computers and Accessibility},
    pages = {125–132},
    numpages = {8},
    location = {Portland, Oregon, USA},
    series = {Assets '06}
}

@inproceedings{Seah2026Rememo,
    author = {Seah, Celeste and Lee, Yoke Chuan and Lee, Jung-Joo and Yen, Ching Chiuan and Zheng, Clement},
    title = {Rememo: A Research-through-Design Inquiry Towards an AI-in-the-loop Therapist’s Tool for Dementia Reminiscence},
    year = {2026},
    isbn = {9798400722783},
    publisher = {Association for Computing Machinery},
    address = {New York, NY, USA},
    url = {https://doi.org/10.1145/3772318.3790461},
    doi = {10.1145/3772318.3790461},
    booktitle = {Proceedings of the 2026 CHI Conference on Human Factors in Computing Systems},
    articleno = {236},
    numpages = {20},
    location = {},
    series = {CHI '26}
}

@article{Boyd2021,
    title={Digital reminiscence app co-created by people living with dementia and carers: Usability and eye gaze analysis},
    author={Boyd, Kyle and Bond, Raymond and Ryan, Assumpta and Goode, Deborah and Mulvenna, Maurice},
    journal={Health Expectations},
    volume={24},
    number={4},
    pages={1207--1219},
    year={2021},
    publisher={Wiley Online Library},
    doi={10.1111/hex.13251}
}

@online{InspireD2026,
    author = {Ulster University},
    title = {InspireD - Reminiscence App},
    year = {n.d.},
    url = {https://www.theinspiredapp.com},
    note = {Accessed on July 8, 2026}
}

@article{Nebot2022LONGREMI,
    title={LONG-REMI: an AI-based technological application to promote healthy mental longevity grounded in reminiscence therapy},
    author={Nebot, {\`A}ngela and Dom{\`e}nech, Sara and Albino-Pires, Nat{\'a}lia and Mugica, Francisco and Benali, Anass and Porta, X{\`e}nia and Nebot, Oriol and Santos, Pedro M},
    journal={International journal of environmental research and public health},
    volume={19},
    number={10},
    pages={5997},
    year={2022},
    publisher={MDPI},
    doi={10.3390/ijerph19105997}
}

@article{Wang2024GoodTimes,
    title={Promoting personalized reminiscence among cognitively intact older adults through an AI-driven interactive multimodal photo album: Development and usability study},
    author={Wang, Xin and Li, Juan and Liang, Tianyi and Hasan, Wordh Ul and Zaman, Kimia Tuz and Du, Yang and Xie, Bo and Tao, Cui and others},
    journal={JMIR aging},
    volume={7},
    number={1},
    pages={e49415},
    year={2024},
    publisher={JMIR Publications Inc., Toronto, Canada}
}

@article{Astell2010,
    title={Stimulating people with dementia to reminisce using personal and generic photographs},
    author={Astell, Arlene J and Ellis, Maggie P and Alm, Norman and Dye, Richard and Gowans, Gary},
    journal={International Journal of Computers in Healthcare},
    volume={1},
    number={2},
    pages={177--198},
    year={2010},
    publisher={Inderscience Publishers}
}

@article{Altizer2025,
    title={The dementia care workforce: Essential to care but large research gaps exist},
    author={Travers Altizer, Jasmine L and Reckrey, Jennifer M and Frogner, Bianca K and Grabowski, David C and Spetz, Joanne},
    journal={Alzheimer's \& Dementia},
    volume={21},
    number={5},
    pages={e70269},
    year={2025},
    publisher={Wiley Online Library}
}

@article{Pond2026,
    title={Evaluating dementia training programs for home care workers: a scoping review},
    author={Pond, Brittney and Neri, Melinda and Suarez Vargas, Kattia and Yeh, Jarmin},
    journal={The Gerontologist},
    volume={66},
    number={3},
    pages={gnaf311},
    year={2026},
    publisher={Oxford University Press},
    doi={10.1093/geront/gnaf311}
}

@online{RT2026,
    author = {Alzheimer's Association},
    title = {Reminiscence and Reminiscence Therapy},
    year = {2026},
    url = {https://www.alz.org/help-support/caregiving/daily-care/reminiscence-and-reminiscence-therapy},
    note = {Accessed on July 1, 2026}
}

@online{Remme,
    author = {Alzheimer's Association},
    title = {Remme - AI-Powered Reminiscence Therapy},
    year = {2026},
    url = {https://tryremme.com},
    note = {Accessed on July 1, 2026}
}

@online{langgraph,
    author = {LangChain},
    title = {LangGraph},
    year = {n.d.},
    url = {https://www.langchain.com/langgraph},
    note = {Accessed on July 1, 2026}
}

@inproceedings{Zhang2025RemiHaven,
    author = {Zhang, Xuechen and He, Changyang and Zhang, Peng and Gu, Hansu and Gu, Ning and Shen, Qi and Hu, Zhan and Lu, Tun},
    title = {RemiHaven: Integrating ``In-Town" and ``Out-of-Town" Peers to Provide Personalized Reminiscence Support for Older Drifters},
    year = {2025},
    isbn = {9798400713941},
    publisher = {Association for Computing Machinery},
    address = {New York, NY, USA},
    url = {https://doi.org/10.1145/3706598.3714277},
    doi = {10.1145/3706598.3714277},
    booktitle = {Proceedings of the 2025 CHI Conference on Human Factors in Computing Systems},
    articleno = {1034},
    numpages = {20},
    series = {CHI '25}
}

@inproceedings{Zhu2026Reminiscope,
    author = {Zhu, Lisha and Qi, Rui and Huang, Siyuan and Li, Xueliang},
    title = {Remembering with Reminiscope: Codesigning with Generative AI for Reminiscence Among Older Adults},
    year = {2026},
    isbn = {9798400722783},
    publisher = {Association for Computing Machinery},
    address = {New York, NY, USA},
    url = {https://doi.org/10.1145/3772318.3791390},
    doi = {10.1145/3772318.3791390},
    booktitle = {Proceedings of the 2026 CHI Conference on Human Factors in Computing Systems},
    articleno = {473},
    numpages = {21},
    location = {},
    series = {CHI '26}
}

@inproceedings{Sun2025ReminiBuddy,
    author = {Sun, Jingwei and Zhang, Zhongyue and Wang, Mengyang and Li, Nianlong and Lu, Zhangwei and Xiang, Yan and Zhang, Liuxin and Zhang, Yu and Wang, Qianying and Fan, Mingming},
    title = {Chorus of the Past: Toward Designing a Multi-agent Conversational Reminiscence System with Digital Artifacts for Older Adults},
    year = {2025},
    isbn = {9798400713941},
    publisher = {Association for Computing Machinery},
    address = {New York, NY, USA},
    url = {https://doi.org/10.1145/3706598.3713810},
    doi = {10.1145/3706598.3713810},
    booktitle = {Proceedings of the 2025 CHI Conference on Human Factors in Computing Systems},
    articleno = {1031},
    numpages = {22},
    series = {CHI '25}
}

@inproceedings{Kim2025RoboBlocks,
    author = {Kim, Callie Y. and Sato, Arissa J. and White, Nathan Thomas and Ho, Hui-Ru and Lee, Christine P. and Hwang, Yuna and Mutlu, Bilge},
    title = {Bridging Generations using AI-Supported Co-Creative Activities},
    year = {2025},
    isbn = {9798400713941},
    publisher = {Association for Computing Machinery},
    address = {New York, NY, USA},
    url = {https://doi.org/10.1145/3706598.3713718},
    doi = {10.1145/3706598.3713718},
    booktitle = {Proceedings of the 2025 CHI Conference on Human Factors in Computing Systems},
    articleno = {1077},
    numpages = {15},
    series = {CHI '25}
}

@online{GeminiVeo,
    author = {Google},
    title = {Gemini Veo 3.1},
    year = {n.d.},
    url = {https://aistudio.google.com/models/veo},
    note = {Accessed on July 1, 2026}
}

\end{document}